\documentclass[letterpaper]{article} 
\usepackage{aaai25}   
\usepackage{times}  
\usepackage{helvet}  
\usepackage{courier}  
\usepackage[hyphens]{url}  
\usepackage{graphicx} 
\usepackage{natbib}  
\usepackage{caption} 
\usepackage{algorithm}
\usepackage{algorithmic}
\usepackage[table,xcdraw]{xcolor}
\usepackage{dirtytalk}
\usepackage{adjustbox}
\usepackage{booktabs}
\usepackage{soul}
\usepackage{listings}
\DeclareCaptionStyle{ruled}{labelfont=normalfont,labelsep=colon,strut=off} 
\floatstyle{ruled}
\newfloat{listing}{tb}{lst}{}
\floatname{listing}{Listing}
\title{Towards Experience-Centered AI: A Framework for Integrating Lived Experience in Design and Development}

\author {
     Sanjana Gautam\textsuperscript{\rm 1},
     Mohit Chandra\textsuperscript{\rm 2},
     Ankolika De\textsuperscript{\rm 3},
     Tatiana Chakravorti\textsuperscript{\rm 3},\\
     Girik Malik\textsuperscript{\rm 4},
     Munmun~De~Choudhury\textsuperscript{\rm 2}
 }
\affiliations {
     \textsuperscript{\rm 1}University of Texas at Austin,
     \textsuperscript{\rm 2}Georgia Institute of Technology,
     \textsuperscript{\rm 3}Pennsylvania State University,
     \textsuperscript{\rm 4}Northeastern University\\
     sanjana.gautam@utexas.edu, mchandra9@gatech.edu, apd5873@psu.edu,   tfc5416@psu.edu, \\malik.gi@northeastern.edu, mchoudhu@cc.gatech.edu
}

\usepackage{bibentry}

\begin{document}

\maketitle

\begin{abstract}
Lived experiences fundamentally shape how individuals interact with AI systems, influencing perceptions of safety, trust, and usability. While prior research has focused on developing techniques to emulate human preferences, and proposed taxonomies to categorize risks (such as psychological harms and algorithmic biases), these efforts have provided limited systematic understanding of lived human experiences or actionable strategies for embedding them meaningfully into the AI development life-cycle. This work proposes a framework for \textit{meaningfully} integrating lived experience into the design and evaluation of AI systems. We synthesize interdisciplinary literature across lived experience philosophy, human-centered design, and human-AI interaction, arguing that centering lived experience can lead to models that more accurately reflect the retrospective, emotional, and contextual dimensions of human cognition. Drawing from a wide body of work across psychology, education, healthcare, and social policy, we present a targeted taxonomy of lived experiences with specific applicability to AI systems. To ground our framework, we examine three application domains— (i) education, (ii) healthcare, and (iii) cultural alignment—illustrating how lived experience informs user goals, system expectations, and ethical considerations in each context. We further incorporate insights from AI system operators and human-AI partnerships to highlight challenges in responsibility allocation, mental model calibration, and long-term system adaptation. We conclude with actionable recommendations for developing experience-centered AI systems that are not only technically robust but also empathetic, context-aware, and aligned with human realities. This work offers a foundation for future research that bridges technical development with the lived experiences of those impacted by AI systems.
\end{abstract}

%

\section{Introduction}

Lived experience signifies the \textit{the view from inside} \cite{casey2023lived} in that it often captures the subjects' interpretation of the object. Lived experience differs based on whether reality is perceived as distinctive values or continuously meaningful in the relation of \textit{object} and \textit{subject}. Contrasting this, within our context of human-computer interaction (HCI) literature, experience is postulated as authentic and having universal qualities \cite{kruks2014women}. The positioning of \textit{experience} as \textit{universal} creates the need for evidence-based truth or belief of self-evidence to be reliable. This can be harmful as evidence-based truth captures singular realities while lived experiences are multifaceted in nature. Thus, it becomes important to contextualize, structure and define lived-experiences. A framework evaluating the role of human lived experiences in AI would provide a comprehensive method for incorporating these experiences into AI design. Examining lived experiences can inform strategies for creating AI agents that set clearer expectations, provide empathetic responses, and adapt to user contexts. For example, AI systems should be designed to probe user goals and intents to personalize interactions \textit{meaningfully}.

\begin{table*}[t]
\small
\centering
\begin{adjustbox}{width=\linewidth}
\begin{tabular}{|>{\raggedright\arraybackslash}m{0.25\textwidth} 
                |>{\raggedright\arraybackslash}m{0.20\textwidth} 
                |>{\raggedright\arraybackslash}m{0.20\textwidth} 
                |>{\raggedright\arraybackslash}m{0.20\textwidth} 
                |>{\raggedright\arraybackslash}m{0.20\textwidth}|}
\rowcolor[HTML]{C0C0C0}
\hline
\textbf{Framework/Paper} & \textbf{Core Approach} & \textbf{Value Basis} & \textbf{Methodology} & \textbf{Unique Features} \\
\hline
Fundamental Value Alignment~\cite{shen2024valuecompass} & Psychological value taxonomy & Schwartz Theory, 49 values & Surveys, value measurement & Systematic value checklist, context-aware scenarios \\
\hline
AI Policy Alignment~\cite{wef2024} & Policy \& guidelines & Societal/shared values & Framework review & Global, policy-oriented \\
\hline
Measuring Human-AI Alignment~\cite{norhashim2024measuring} & Empirical measurement & Human values (surveyed) & Quantitative analysis & Focus on LLMs, misalignment quantification \\
\hline
Bi-Directional Alignment~\cite{shen2024towards} & Human-centered, bidirectional & Human/AI adaptation & Systematic review & Mutual adaptation emphasis \\
\hline
Human Well-being~\cite{van2023framework} & Wellbeing, cybernetic theory & Eudaimonic wellbeing & Human-centered design & Community-led, feedback loops \\
\hline \rowcolor{green!20} 
\textbf{Our Paper} (LEAF) & Lived experience centered & Contextual, emotional, retrospective & Interdisciplinary synthesis, domain-specific case studies & Embeds lived experience in lifecycle, empathy, responsibility allocation \\
\hline
\end{tabular}
\end{adjustbox}
\caption{\textbf{Comprehensive Overview of Human-AI Value Alignment Frameworks}, including a comparison of our lived experience-centered approach with prior work.}
\label{table_prev}
\end{table*}

Human-AI value alignment has been a focal point of ongoing and widespread deliberation at both AI and HCI venues \cite{shen2024valuecompass, shen2024towards, turchin2019ai, terry2023interactive, van2023framework, chandra2024lived}. The diversity of human values across cultures presents challenge for achieving meaningful AI alignment. Prior work has covered critical discourse and  provides theoretical frameworks for value alignment in the context of generative AI in the workforce, with a focus on community and organizational values \cite{shen2024valuecompass}. Aligning AI systems with nuanced, context-specific human experiences can lead to more equitable, inclusive, and effective outcomes in human-AI interaction \cite{terry2023interactive}. Unlike existing approaches that rely heavily on psychological taxonomies or quantitative surveys, this work emphasizes the integration of lived experiences throughout the AI development life cycle, not as a static checklist but as a dynamic and continuous consideration \cite{van2023framework, wef2024}. Our paper addresses key limitations in prior alignment frameworks by centering lived experience — defined as retrospective, emotional, and contextual— as foundational to AI system design and evaluation (Table \ref{table_prev}). Drawing from literature in philosophy, human-centered design, and HCI, \textbf{the LEAF framework} (\textbf{L}ived \textbf{E}xperience Centered \textbf{A}I \textbf{F}ramework) situates lived experience as critical to understanding user interaction, trust, and well-being, particularly in domain-specific contexts such as education, healthcare, and cultural alignment. By doing so, it highlights the need for empathetic, context-aware AI systems that reflect and respond to the realities of those they impact. 

A systematic exploration of lived experiences in AI also necessitates methodological pluralism. A truly human-centered AI approach must consider subjective experiences, embodied intelligence, and first-person perspectives \cite{bingley2023human, capel2023human}. This involves recognizing the primacy of lived experience in understanding human cognition and integrating multi-modal sensory communication into AI systems \cite{rix2024understanding, dieumegard2021lived}. Memory and retrospective experience play a critical role in how humans navigate the world, and AI models should strive to mirror these aspects more effectively \cite{howard2018memory, medicalxpressPerceptionWorking, chiorri2024subjective, kahneman2005living}.  In this work, we present \textbf{the LEAF} that emphasizes the integration of human lived experiences throughout the AI development life cycle, guiding their incorporation into both design and evaluation processes. 

This paper positions lived experience as a foundational lens for aligning AI systems with the nuanced realities of human users, focusing on design, evaluation, and deployment processes. We first define lived experience and explore its relevance in shaping human–AI interactions, especially in high-stakes and everyday contexts. We then introduce a conceptual framework structured around four key dimensions of lived experience: (i) sense of self, (ii) health, (iii) social and cultural identity, and (iv) learning. Building on this framework, we identify specific stages within the AI development pipeline where lived experience can be meaningfully integrated and offer actionable design recommendations to support the inclusion of diverse experiential perspectives. Through real-world case studies in domains such as education, healthcare, and religion, we demonstrate how attention to lived experience can inform more inclusive, empathetic, and context-aware AI systems.  Thus, our work contributes a structured approach for embedding lived experience into the lifecycle of AI systems, with the goal of guiding more responsible and socially grounded AI practice.

\section{Facets of Lived Experiences}

The concept of lived experiences goes back as late nineteenth century \cite{casey2023lived}. Although scientific research has shifted over time—from being primarily observational in the 18th and 19th centuries to becoming more experimental and data-driven today, there is a renewed recognition of the value of lived experience. It can be traced, historically, to documentation of personal or observational accounts have led to both societal and scientific progress \cite{scott1991evidence}. This section explicates the definitions, dimensions and importance of lived experience relevant to the LEAF framework.


\subsection{\label{tax}What Do we mean by Lived Experiences - Taxonomy of Lived Experiences} 

The Oxford English Dictionary defines \say{\textit{\textbf{Lived Experience}}} as``\emph{Personal knowledge about the world gained through direct, first-hand involvement in everyday events rather than through representations constructed by other people. It may also refer to knowledge of people gained from direct face-to-face interaction rather than through a technological medium}''\cite{oxf}. Likewise, the concept of lived experience has been studied across disciplines such as psychology, philosophy, sociology, and clinical science, serving as an important lens to deepen our understanding of human experiences and realities. The term \emph{lived experience} was first popularized within the field of philosophy with the lens of objectivity~\cite{bunnin2008blackwell, casey2023lived}, particularly through the work of Wilhelm Dilthey, to represent individual's embodied, ﬁrst-person, pre-reﬂective experience of themselves and their world. Within Philosophy, Phenomenology specifically focused on lived experiences and aimed to understand the nature of conscious experiences and the significance of phenomena from a first-person perspective~\cite{sep-phenomenology}.

 As articulated in Heidegger’s philosophy \cite{heidegger1962being}, lived experience is rooted in the ontological structure of human existence, conceptualized through the notion of Dasein, or \say{being-in-the-world}. Rather than viewing individuals as isolated observers, Heidegger emphasizes that human beings are always already situated within a particular social, historical, and relational context \cite{korteling2021human}. Lived experience, in this view, is shaped through the continuous and intimate interaction between self and world \cite{miles2013exploring}. It is not merely about what is observed, but how being is experienced from within—contextually, temporally, and meaningfully \cite{korteling2021human}. Hermeneutic phenomenology, drawn from Heidegger’s ideas, therefore seeks to uncover the implicit meanings embedded in everyday life, interpreting these meanings as expressions of one’s existence in the world \cite{spiegelberg2012phenomenological}.

Within sociology, the concept of lived experience encompasses individual perceptions, interpretations, and emotional responses to life events that are often shaped by social, cultural and historical factors involving interactions with others~\cite{docmckeeLivedExperiences}. Past research has explored various facets of lived experience ranging from personal experiences, work-related experiences, and social relationships. Below, we highlight examples from AI ethics research and related fields that illustrate how lived experiences have been centered in the design and development of AI systems.

 \begin{figure}
   \centering
   \includegraphics[width=0.75\linewidth]{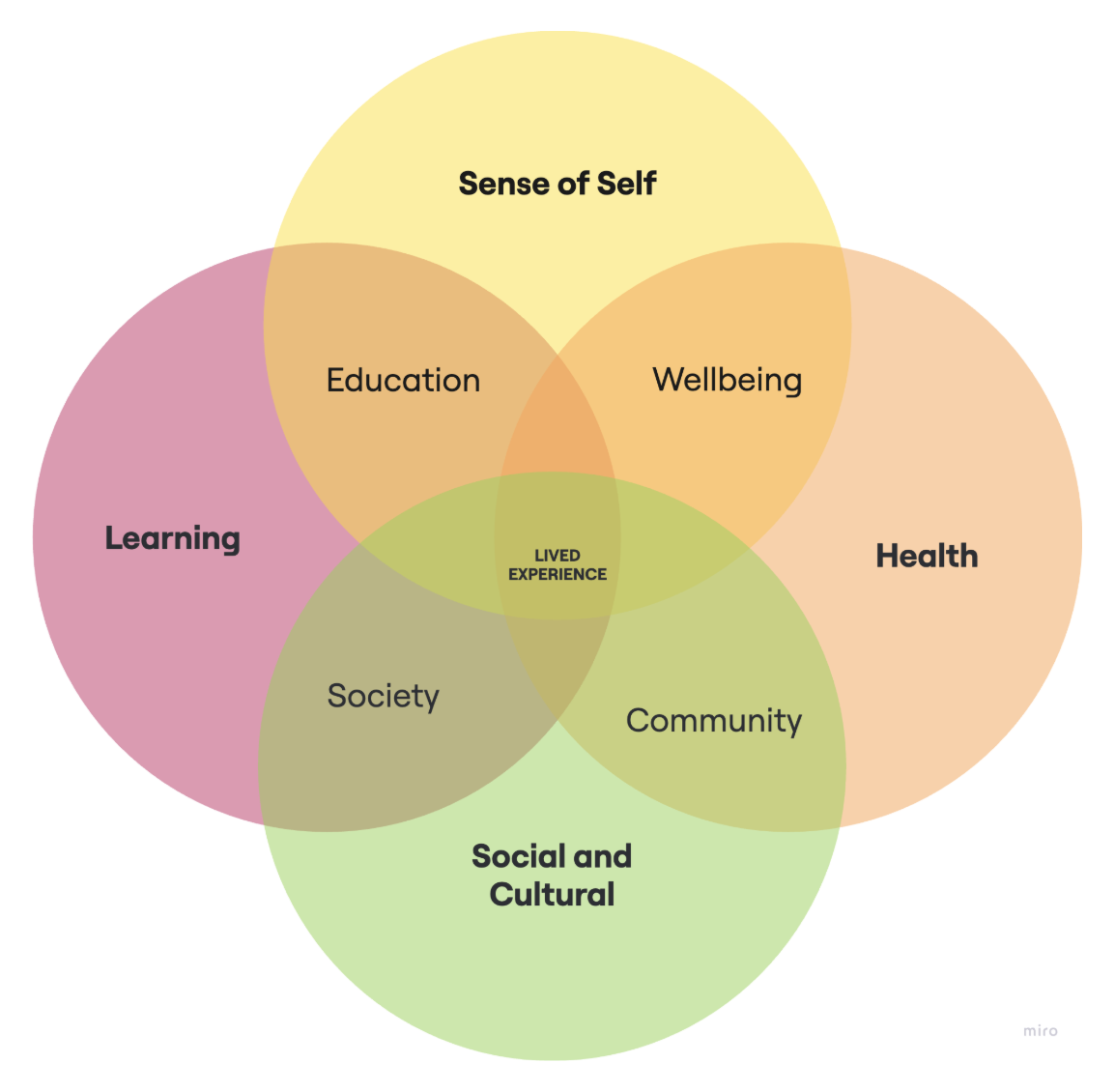}
   \caption{Dimensions of lived experience drawn from cross-disciplinary literature in psychology, artificial intelligence, healthcare, and human-computer interaction. These dimensions serve as interpretive guides for situating lived experience within sociotechnical systems.}
   \label{fig:dim}
 \end{figure}

~\citet{portway2005you} conducted a study understanding experiences of young adults with Asperger's syndrome within the society. In another study, researchers focused on understanding the association between well-being, sense of belonging, connectedness to community ~\cite{haim2023sense}. They found that sense of belonging was associated with increased well-being. Feminist theory has been a prominent framework for examining how gender and social relations shape the lived experiences of women and other marginalized gender groups~\cite{garko1999existential,patterson2016black}. In addition to gender identity, cultural background has also been recognized as an important factor in shaping individual experiences~\cite{pufall2010navigating}.

In psychology and clinical science, lived experience has been recognized as a key component towards understanding patient well-being. Past works have explored ways to integrate lived-experience perspectives of patients in mental healthcare~ \cite{happell2007consumer,walsh2009improving}. Past research has also shown that integrating lived experiences can lead to higher-quality research outcomes, better acceptance of treatment procedures and increased patient outcomes~\cite{brett2014mapping, goodare1999involving, beames2021new}. Other works have explored ways to integrate patient's lived experience into the clinical process pipeline. For instance,~\citet{otado2015culturally} examined strategies for overcoming barriers towards recruiting individuals from African American communities. Such strategies included providing informational sessions and disseminating newsletters about study outcomes.

Lived experience has also been valued in educational settings for both educators and learners. Past works have shown that integrating lived experience of end consumers within health profession education leads to more empathetic and nuanced understanding of health and illness among health students and professionals~\cite{soon2020consumer}.~\citet{matu2024importance} found that implementing culturally relevant pedagogy that encompassed diverse lived experiences fosters a culturally diverse learning environment, leading to enhanced cultural competence, critical thinking, global citizenship and academic achievement.

Lived experiences have also been explicated as \textit{expertise}, with international consensus, especially in health, design and technologies \cite{muchamore2024lived, chandra2024lived, chandra-etal-2025-lived}. Likewise, scholars in personal informatics have emphasized the importance of incorporating lived experiences and subjectivity into data-centric systems that, as they note, \textit{``attend to people’s subjective perspectives on personal informatics}" \cite{cosley2017lived}. This comes from the idea that, the focus on rational self-improvement in personal informatics often overlooks the lived, emotional, and evolving experiences people have with tracking technologies, which can exclude some users and even cause harm when misaligned with their actual needs \cite{10.1145/2556288.2557039, cosley2017lived}.

While the concept of lived experience has been extensively examined within the aforementioned disciplines, it remains widely misunderstood and largely overlooked in AI research, particularly within natural language processing \cite{girju2023understanding}. Past works have also highlighted the distinction between human-intelligence from artificial intelligence (AI), making it important to understand the role of lived experience within AI~\cite{olivier2017artificial}. Hence, in this work, we examine what constitutes the concept of lived experience specifically in the context of AI, and how it can inform the development and design of more user-centered AI systems. For the purposes of this framework, we define \textit{"user"} to include both the end consumer of the technology and the facilitator who leverages it. For example, in the case of an AI chatbot used in a healthcare setting, both the care provider using the chatbot and the patient interacting with it are considered users.

\subsection{Importance of Lived Experience Within Technology Design}

AI development has traditionally treated users as abstract data subjects, reducing their behavior to patterns that can be learned and optimized. As~\citet{doi:10.1177/20539517211061122} argue, the lived experience of data subjects remains deeply embedded in their life worlds, and cannot be fully understood through data alone \footnote{https://www.mhc.wa.gov.au/our-initiatives/our-projects/lived-experience-(peer)-workforce-project}. If AI systems are to be trustworthy, fair, and responsive to the people who use them and more importantly the people they affect, their development must incorporate not only technical metrics but also the grounded, subjective knowledge that individuals bring \cite{freiman2023trustworthy, 516138}. Thus, designing with lived experience does not replace data-driven methods \cite{markham2019experimenting}, but complements them as they have the potential of exposing blind spots \cite{frank2024blind}, surfacing harms \cite{10.1145/3274424}, and foregrounding the perspectives of those often marginalized in the design process. It is important to note here that while marginalized communities face a higher risk of erasure \cite{cantley2025indigenous}, this framework is guided towards all communities. Mentions of marginalized communities, henceforth, are as an illustrative example that should extend to other communities as well. Likewise, AI ethicists, have often explicated, that lived experiences be accounted for when building systems, and have attempted to do so, especially in systems and technologies in social and public sectors (See for example, \cite{10.1145/3531146.3533157}). 

In recent times, as AI is increasingly repurposed and appropriated for knowledge generation and sharing, lived experience has gained importance due to the shifting epistemologies introduced by technologies such as large language models \cite{mugleston2025epistemology}. As these systems increasingly mediate knowledge production and dissemination, they risk flattening culturally specific ways of knowing \cite{amershi2020culture}. Likewise, it is important to note that knowledge is not merely abstract or universal -- it is often rooted in lived, situated experience. Expertise emerges through embodied engagement with the world, not just formal instruction \cite{casey2023lived}. AI systems that ignore these dynamics, risk distorting or disregarding the very contexts that lend meaning to knowledge.

We now turn to a more in-depth examination of how these conceptual differences manifest in the design of artificially intelligent systems. A recurring critique of lived experience as a construct is its perceived lack of scientific rigor and its limited applicability to empirical, real-world contexts \cite{sssmaggi}. This critique highlights a broader epistemological divide between experiential knowledge and traditional scientific paradigms \cite{allison2000shall}. In this paper, we aim position ourselves within that gap, also attempting to bridge it by situating lived experience within a scientifically grounded framework. Given the inherently subjective and deeply embedded nature of these issues within human experience, we argue that integrating lived experiences into the design and evaluation of AI systems can offer critical insights and serve as a valuable strategy for mitigating these harms.


\subsection{Dimensions of Lived Experiences}

Building on prior research examining bias and harm in AI systems \cite{shelby2023sociotechnical}, we offer a lens for organizing and understanding how lived experiences manifest in relation to AI design and use. Figure~\ref{fig:dim} presents a visual presentation of the organization. Rather than presenting rigid or discrete categories, we identify seven interrelated thematic dimensions that commonly surface in the literature. These dimensions—drawn from cross-disciplinary frameworks in psychology, AI, healthcare, and HCI, serve as interpretive guides for situating lived experience within sociotechnical systems. In the sections that follow, we outline each dimension with illustrative examples, recognizing that they are fluid, overlapping, and context-dependent rather than mutually exclusive or exhaustive.

\begin{itemize}
    \item \textbf{Sense of Self} : Among the most difficult dimensions of lived experience to quantify are those rooted in personal, introspective reflection \cite{schwarz1996feelings}. These experiences often inform individual consumption patterns and preferences, shaped significantly by reflective processes. Reflective experiences capture the retrospective interpretation of events, rather than their immediate, in-the-moment perception \cite{kahneman2005living}. Such reflection frequently leads to reflective learning, which plays a crucial role in shaping how lived experiences evolve for individuals \cite{boud2013reflection}. Intuition and tacit knowledge are central to this process, aligning with Schön’s theory of reflective practice, which emphasizes learning through thoughtful engagement with one's actions and experiences \cite{schon1979reflective}. Closely tied to reflective experience is the concept of metacognition—the awareness and regulation of one’s cognitive processes \cite{schraw1994assessing}. Predominantly studied within learning sciences, metacognitive awareness enables individuals to evaluate and adapt their learning strategies \cite{wade1989developing}. Interaction with AI systems, particularly those used in educational settings, can disrupt existing mental models of learning by introducing alternative paths of information navigation and knowledge construction \cite{do2024evaluating}. Emerging research on the use of AI in learning contexts illustrates how these systems influence metacognitive patterns and learning behaviors \cite{fan2025beware}. When considering lived experiences that inform one’s sense of self, it is important to account for the complex and often deeply personal influence of prior traumatic experiences. These histories can shape emotional responses and engagement with AI technologies in ways that are not easily predicted or generalized. Such instances require nuanced, context-sensitive approaches that extend beyond standard user experience models and design practices \cite{siddals2024happened}. 
    \item \textbf{Health} : Contexts in health are heavily impacted by lived experiences. As conversational agents become more common in mental health contexts—often valued for the perceived privacy, nonjudgmental interactions, and emotional safety they offer users \cite{lee2025artificial}-- it is essential to integrate the lived experiences of health practitioners into the design and governance of such systems. These practitioners possess expertise through years of experiences in navigating complex emotional scenarios, which current AI systems often lack. Research has shown that users may turn to general-purpose AI during moments of psychological distress \cite{song2024typing}, sometimes appreciating the model’s perceived neutrality, even when cultural awareness is absent. However, such use cases can be unpredictable and highlight the limitations of designing for narrowly anticipated needs. Health practitioners’ experiential knowledge is therefore crucial not only for identifying risk but also for understanding edge cases and contradictions that may otherwise go unrecognized. Their involvement can help ensure that AI tools are contextually appropriate and safe, especially in high-stakes scenarios \cite{olawade2024enhancing}. In medical consultation, rising healthcare costs and limited access to care have fueled reliance on AI tools. Even before the AI boom, platforms like WebMD reflected a public need for alternative guidance \cite{mobihealthnewsInDepthWebMD}. Today, conversational medical agents continue this trend, but often fail to reflect the nuanced knowledge of practitioners. For instance, LLMs struggle to address adverse psychiatric drug reactions and rarely account for trauma or disability contexts \cite{chandra-etal-2025-lived}. Without grounding in lived practitioner and patient experience, such systems risk offering decontextualized and potentially harmful advice.
    \item \textbf{Social and Cultural} : We intentionally address social and cultural contexts together to emphasize their intertwined nature across most lived experience scenarios. This entanglement often lends cultural experiences a perceived authenticity—one that may be accepted uncritically rather than subjected to reflective scrutiny. Consequently, when researchers draw on individuals’ experiences, there is a risk of mistaking these accounts as direct, unmediated sources of truth, rather than recognizing them as shaped by complex socio-cultural frameworks \cite{mcintosh2019exploring}. As introduced in the section above, foundational feminist theorists such as Simone de Beauvoir \cite{kruks1992gender} and Edith Stein \cite{feldhay1994edith} emphasized the importance of recognizing sexual difference as integral to women’s embodied experience and consciousness. Beauvoir, in particular, drew on Husserl’s distinction between the body as an object of detached observation (Körper) and the body as it is subjectively lived and experienced (Leib) \cite{coolen2014bodily}. Thus, gender identities and lived experiences are deeply influenced by social, geographic, and political contexts, which in turn shape how individuals engage with AI systems \cite{armutat2024artificial}. These contextual differences contribute to varied human behaviors and perceptions, especially regarding technology adoption and trust. Likewise, this becomes an important example of how social and cultural nuances of lived experience have significant implications, affecting outcomes such as workplace hiring, representation in socio-technical domains, and the equitable deployment of AI technologies \cite{an2024large, o2024gender}.
    \item \textbf{Learning} : One of our greatest human capacities is the ability to learn from both personal and others' experiences. Yet this form of knowledge is often undervalued, as rational prioritizing external sources of information over lived human experience \cite{farrell2020researching}. Students’ mental models of learning, informed by both personal experience and educational theory \cite{jonassen1999mental}, shape how they engage with instruction. Replacing human educators with AI systems risks overlooking the contextual expertise instructors develop over years of practice \cite{luckin2016intelligence}. Much learning occurs beyond formal instruction—for example, through peer tutoring fostered by classroom relationships \cite{selwyn2019should}. For example, in Giles’ study \cite{giles2012exploring}, grounded in Heideggerian and Gadamerian philosophy \cite{magrini2011recovering}, the authors show that teacher–student relationships are foundational to the educational experience yet often overlooked.Additionally, instructors contribute by integrating diverse student backgrounds and building classroom resilience, offering a depth of lived experience that current AI systems struggle to replicate.
\end{itemize}

 \begin{figure*}
   \centering
   \includegraphics[width=1.0\linewidth]{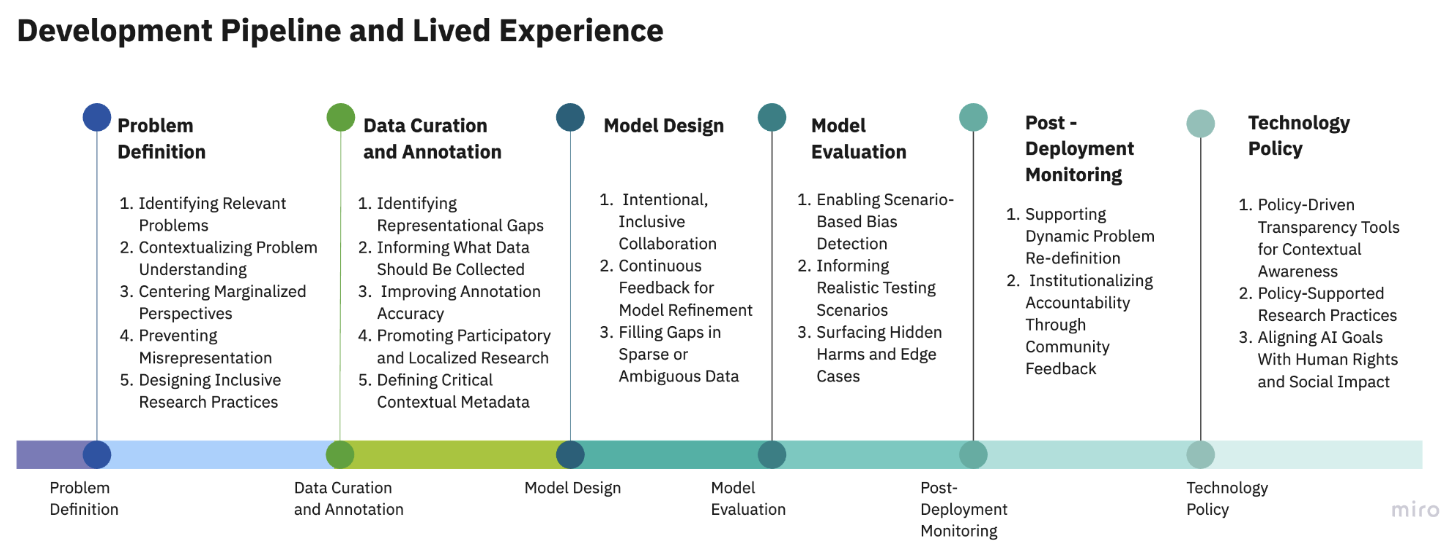}
   \caption{Outline of a typical AI development pipeline and how the LEAF can be leveraged to incorporate the human lived experience at each stage. Each stage name is highlighted \textbf{bold} (eg: Model Design), followed by processes recommended in the LEAF to meaningfully incorporating human lived experiences.}
   \label{fig:mapp}
 \end{figure*}

Given the discussion above, it is also important to talk about dimensions that are complex. As previously discussed, lived experiences play a crucial role in facilitating this alignment. A key facet of both cultural context and personal identity is moral alignment \cite{naous2023having}. While substantial work has been undertaken to establish alignment with human values \cite{han2022aligning, frempong2024harmonizing}—resulting in the development of alignment scales and frameworks—moral alignment remains particularly challenging. This difficulty stems from the divergent and often conflicting values that arise from personal, community, cultural, national, and other contextual layers \cite{markus1999conflictways}. Such complexity resists straightforward categorization, as moral orientations are not easily captured by static typologies. In this regard, lived experiences offer critical insights that can inform and enrich the design of AI systems. 

\section{Lived Experience Framework within the AI Development Pipeline}


Incorporating lived experiences into AI development workflows presents different opportunities to rethink the design techniques where technology is used for human good, especially in computer science, biomedical, humanities, and social sciences. We can include lived experience at any stage of designing, developing, and deploying AI systems, but its role and impact differ depending on the phase and the application. To this end we propose \textbf{Lived Experience Centered AI Framework (LEAF)}. It ensures that AI models are grounded in real-world concerns, especially those of communities often marginalized in technical design processes. This creates the possibility for co-creative design, where tools reflect technical feasibility and cultural and social relevance. In this section, we have explained how we can integrate lived experiences throughout the research life cycle. Figure~\ref{fig:mapp} presents a visual representation of the LEAF.

\subsection{Problem Definition}

Lived experiences can play a critical role in shaping the early stages of AI development, particularly in defining the problem space. Rather than relying on technical abstractions or dominant normative frameworks, incorporating lived experience brings attention to how people actually encounter and are affected by a given issue. This perspective helps uncover situational nuances that are often overlooked in top-down design approaches \cite{ajjawi2024researching}. The ``gulf of execution'', as described by Norman \cite{norman1988psychology}, highlights the disconnect between a user's intentions and the system’s affordances or interface. This gulf can be significantly widened when systems are not designed with attention to users’ contextual realities. Lived experience—encompassing people’s tacit knowledge \cite{mascitelli2000experience}, cultural contexts \cite{ginn2016promoting}, and everyday practices \cite{ajjawi2024researching} — helps bridge this gap by aligning the psychological language of users with the operational language of the system. When excluded from the early design process, systems risk embedding the assumptions of developers rather than reflecting the diverse needs of actual users, especially those from marginalized communities.

For instance, in natural language processing tools aimed at detecting depression or anxiety on social media, researchers have found that standard models often misclassify language used by LGBTQ+ or neurodivergent users, whose expressions of distress deviate from dominant linguistic norms \cite{may2019measuring}. Similarly, predictive policing tools such as PredPol, developed without engagement from overpoliced communities, reproduced systemic bias and flawed assumptions about criminal behavior \cite{mugari2021predictive}. These failures underscore the importance of integrating lived experience at the problem-definition stage. Participatory approaches—such as co-design workshops or speculative design sessions—can facilitate this integration, grounding early prototypes in the realities of those most affected. For example, Brogden et al. \cite{brogden2024enhancing} reflect on the value of including young individuals with lived mental health experiences in the design of digital health tools, showing how such engagement improves relevance and trust.

\subsection{Data Curation and Annotation}
During data curation, lived experience helps identify representational gaps, and guides more inclusive annotation practices. We can include lived experience to understand what data should be included, how it should be labeled, and what contextual metadata is critical for humanities during the data sourcing, selection, labeling, and validation stages of AI development to mitigate bias. For example, medical datasets are often male-dominated or exclude non-binary individuals, leading to diagnostic tools that fail to detect symptoms in women or gender-diverse patients\cite{obermeyer2019dissecting}. We need a community-led data annotation process because annotators without relevant lived experience may mislabel content, especially when it involves health data, cultural practices, or marginalized identities. Another example is how research in NLP lacks the geographical diversity and participatory research can help to mitigate it. The Masakhane project for African NLP involves native speakers in annotation and validation to preserve linguistic and cultural accuracy\cite{nekoto2020participatory}.

\subsection{Model Design}
Lived experience plays a critical role in aligning AI system design with the social realities and contextual nuances of users. Participatory design \cite{muller1993participatory} and value-sensitive design (VSD) frameworks \cite{umbrello2018value} demonstrate how user narratives — especially accounts of exclusion, discomfort, or misrecognition — can inform concrete design decisions, such as refining system logic, interface affordances, or feedback mechanisms. These approaches help define abstract technical goals like fairness, interpretability, or accuracy in domain-specific and socially meaningful ways. Human-AI collaboration methods, particularly human-in-the-loop systems \cite{sicilia2023humbel, robertson2024human}, operationalize this alignment by incorporating ongoing feedback from community members or domain experts to guide data labeling, decision-making, and model refinement. This iterative engagement ensures that models are not only technically accurate but also culturally relevant, emotionally attuned, and ethically grounded. Far from being at odds with lived experience, human-AI collaboration offers a robust strategy for embedding it throughout the AI development pipeline.

\subsection{Model Evaluation and Testing}
Developers should invite users or stakeholders affected by the AI systems to test it and evaluate its outputs in real-world or simulated scenarios \cite{mckenna2024bringing, beames2021new}. Participatory evaluation by users or community-centric testing can surface harms, usability issues, and unintended effects that are invisible to developers. Koenecke et al.\cite{koenecke2020racial} demonstrated how racial disparities in speech recognition systems became evident when testing involved diverse users, showing that systems performed significantly worse on Black speakers. Scenario-based testing \cite{carrol1999five, rosson2007scenario}, grounded in real-life contexts from stakeholders such as teachers, patients, and gig workers, helps surface edge cases and bias tied to specific identities or environments. Structured workshops with affected communities can further contextualize model errors and unpack their real-world implications.

\subsection{Post-deployment Monitoring}
A feedback loop is needed to be established if we wanted to implement lived experiments into post deployment monitoring. Feedback loops can continuously surface real world usage insights, harms, and unintended consequences. Rather than relying completely on technical performance metrics, systems should incorporate participatory monitoring mechanisms such as community reporting tools, ethnographic studies, or user diaries that allow diverse stakeholders to share how AI impacts them in daily practice. This is especially vital for AI applications, where language, context, and identity intersect in complex ways. A growing body of research emphasizes continuous post-deployment audits rooted in community experiences. For instance, \cite{koenecke2020racial} in “Racial disparities in automated speech recognition” demonstrate how overlooking user variation can lead to discrimination, highlighting the need for real-world evaluations involving diverse populations.

\subsection{Role of Policies in Integration of Human Lived Experiences}

Policies can embed lived experience into AI development through a range of mechanisms across the AI life cycle. Policies can mandate participatory design processes to ensure the inclusion of community experiences during the development of AI systems. For example, Costanza Chock \cite{costanza2020design} argues that marginalized community experiences should be actively involved in defining what problems AI systems solve and how they are implemented. Public-sector procurement policies—such as New York City's Automated Decision Systems Task Force have begun to recommend participatory mechanisms for algorithmic accountability. Policies can focus on making technical documentation that integrates real-world and sociocultural contexts. Mitchell et al. \cite{mitchell2019model} introduced Model Cards, and Gebru et al. \cite{gebru2021datasheets} proposed Datasheets for Datasets, both advocating for the inclusion of intended use cases, user feedback, and potential risks. Countries should implement a governmental AI accountability policy, requiring public agencies to disclose how automated systems may affect rights and freedoms. Raji et al.\cite{raji2020closing} in “Closing the AI Accountability Gap” propose a structured audit framework that includes community feedback as a key accountability mechanism. These formats create transparency and ensure systems are not deployed in ways that contradict the values and lived experiences of their users.

\section{A Closer look at LEAF}

To demonstrate the applicability of our framework across diverse domains, we present four case studies. These examples show how integrating lived experiences during the problem definition and development stages of AI leads to more context-aware, ethical, and socially aligned systems. Each case study should exemplifies one or more aspects of the framework (e.g., participatory design, contextual evaluation, post-deployment monitoring).

\subsection{Case Study 1 : Students and the Autograder}

In this section we use the scenario of AI application within schools.With the growing digitization of educational practices, computers have become integral to educational institutions. Teaching computer science has been a challenge due to the continuously moving goal post. And for the final push recently the advent of large language models has created for a challenging yet opportunistic environment where technology has disrupted somewhat the natural flow of things. Some major challenges for educational AI include misinformation, hallucination, privacy concerns and rising expenses of running such systems \cite{zainuddin2024does}. 

\subsubsection{The Problem}

We take the sample case of a study \cite{li2023wrong} that illustrated how automatic grading can impact learning in students. The study was conducted in a 600-student, online introductory computer science course for non-majors at the University of Illinois in Fall 2021, focused on teaching Python and Excel. The assignment was a Explain in Plain English (EiPE) problem. EiPE rose in prominence as a method to assess students’ ability to read code and discuss its behavior at a high level of abstraction \cite{whalley2006australasian}. The course used a custom autograder with 87\% accuracy to score EiPE problems, and the study examined how its grading errors affected student learning. This autograder was specifically developed for use in the course. The study evaluated how false positives and false negatives impact students' learned outcome. 

\subsubsection{What did the Study Find?}

They used Bayesian hierarchical generalized linear models to analyze participant behavior on practice and post-test questions, predicting learning outcomes, self-assessment accuracy, and feedback engagement time, while applying \textit{weakly informative priors} to ensure reliable inference despite a small sample size. The study found that when the AI grader provided overall positive feedback, students were less likely to read the detailed feedback and more likely to reject any negative or incorrect evaluations it gave. When the main goal is learning, it's important to avoid false positives—but this study shows we need real data to weigh the trade-offs, since intuitive choices may not lead to the best results. So we see that missing student experiences' in the design led to failures in learning outcomes. 

\subsubsection{Role of Lived-experience}

To further understand the points of failure, we revisit the study details through the lens of our LEAF. Formative feedback is a cornerstone of effective learning, and when automated systems misgrade or oversimplify assessment tasks, the educational experience is diminished. This example highlights the need to contextualize problem understanding—recognizing that instructional dynamics are not merely technical, but relational and pedagogical. By centering the perspectives of instructors, tutors, and students,  designers can identify how feedback is interpreted, internalized, and acted upon in diverse learning settings. 

\subsection{Case Study 2 : Conversational Clinical Agents}

Early versions of AI have been used for medical diagnosis~\cite{mckinney2020international} and virtual healthcare assistance~\cite{fitzpatrick2017delivering}. With the advancements in LLMs, AI has the potential to transform the field of healthcare through increasing diagnosis efficiency~\cite{diagnosisnature25, williams2024use, kern2024assessing}, newer scientific discoveries~\cite{theodoris2023transfer, m2024augmenting}, and democratizing access to medical information~\cite{clusmann2023future}. However, integrating AI to real-world use-cases in a high-risk domain such as healthcare requires careful considerations related to AI model's performance and broader implications attached to the use of AI in healthcare. 

\subsubsection{The Problem}

Prior work have increasingly focused on developing and evaluating diagnostic AI systems for single-turn question answering or knowledge retrieval, often benchmarking AI models against medical exams or clinical QA datasets~\cite{singhal2023large, jin2024better, singhal2025toward}. While such approaches have provided insights about the shortcomings of AI models, these often fail to capture the nuances and complexity of real-world experiences of patients healthcare providers \cite{blagec2023benchmark, jacob2025ai}. Additionally, real-world multi-turn clinical conversations require additional evaluation axes such as  empathy, trust-building, relationship-building, respect for the individual and communication efficacy \cite{jacob2025ai, albahri2023systematic}. These dimensions of evaluations are important for supportive healthcare communication. However, developing holistic evaluation frameworks for AI in healthcare remains an open challenge.

\subsubsection{What did the Study Find?}

In response to the challenges outlined above,~\citet{tu2025towards} introduced AMIE (Articulate Medical Intelligence Explorer), a large language model designed for multi-turn diagnostic clinical conversations. They introduced a self-play based simulation environment that took account of the lived-experiences of individuals using vignettes belonging to different demographics background. By capturing the contextual richness of patient encounters, the study foregrounded both clinician and patient perspectives through a dual-metric evaluation framework. This included clinician-centric and patient-centric metrics assessed via automated and human evaluations. Using a combination of automated and human-evaluation, authors observed that AIME outperformed primary care physicians across 28 out of 32 evaluation axes from the specialist physician perspective and 24 out of 26 evaluation axes from the patient actor perspective. 

\subsubsection{Role of Lived Experience}

By centering the lived experiences of both patients and clinicians, this work offers a compelling model for integrating experiential knowledge into the design and evaluation of medical AI systems, thereby enhancing their contextual sensitivity and real-world applicability. This dual-perspective design integrates a multi-centered human layer to the validation of medical AI systems, enabling a more holistic and grounded evaluation of AI's real-world readiness. As illustrated within this example, lived experience is of vital importance within healthcare domain. Past works have highlighted that those directly affected by health issues possess unique insights that can significantly improve healthcare systems, research, policies, and programs~\cite{sartor2023mental}. However, integrating lived experience based insights in AI based systems has been seen as a challenge, where past works have shown the LLMs fail to account for experiences based on race~\cite{yang2024unmasking}, gender~\cite{pfohl2024toolbox}, languages spoken~\cite{jin2024better}, and culture~\cite{omar2025evaluating}. Additionally, past works have also shown that LLMs suffer from lack of experiential knowledge that healthcare providers gain through practice~\cite{chandra-etal-2025-lived}.

\subsection{Case Study 3 : Gods and Machines}

In this section, we examine the importance of cultural alignment in AI systems, focusing on how the integration of cultural values, beliefs, and practices into AI development can impact the effectiveness, integrity and actual use of these systems \cite{bravansky2025rethinking}. As we introduced earlier in this paper that culture is a critical dimension of lived experience. Cultural alignment refers to the extent to which AI technologies consider and respect the cultural contexts in which they operate \cite{gabriel2020artificial}. When AI systems are developed without such alignment, they risk reinforcing biases, causing harm, or overlooking critical cultural nuances \cite{liu2024cultural}. We use the example of AI’s engagement with sacred religious texts to highlight the challenges and implications of cultural misalignment in AI technologies, for specific uses.

\subsubsection{The Problem}

In the context of AI, particularly those that process or interpret culturally significant data (such as religious texts), it is important not to treat the data as neutral or context-free, but embedded within complex historical, cultural, and spiritual frameworks.For example, AI models analyzing sacred texts like the Bible or Quran must respect their religious, cultural, and historical significance beyond just linguistic content. Thus the approach needs to account for how these texts are interwoven with cultural identities, belief systems, and traditions, so as to not create a misalignment between the system’s design and the communities it serves. A key concern centers on AI's capacity to comprehend and reproduce the deeply personal and experiential dimensions of religious life \cite{tampubolon2024artificial}. The lack of cultural alignment in AI can result in systems that fail to respect the spiritual significance of these texts or inadvertently perpetuate cultural erasure \cite{qadri2025risks}.

\subsubsection{What did the Study find?}

Hutchinson’s \cite{hutchinson2024modeling} study highlights how sacred religious texts are frequently used in AI research (for machine translation, corpus creation, and data analysis) without adequate attention to their cultural or spiritual significance. Texts like the Bible and the Quran are often selected for their size, multilingual content, and accessibility, rather than their religious meaning. This reflects a broader tendency in AI to treat data as purely informational, neglecting the embedded cultural and ethical dimensions \cite{tampubolon2024artificial}. Building on this, scholars such as Bostrom \cite{robert2017superintelligence} and Floridi \cite{cath2018artificial} have raised ethical concerns about AI’s role in religious contexts, questioning the authenticity of AI-mediated religious experiences and the appropriateness of delegating spiritual authority to non-sentient systems \cite{tampubolon2024artificial}.

\subsubsection{Role of Lived Experiences}

To prevent the risks of cultural misalignment, Hutchinson  \cite{hutchinson2024modeling} advocated for a more nuanced and culturally sensitive approach to AI development. As highlighted in our framework, AI systems engaging with sacred or culturally significant data must be designed with an awareness that such texts are not neutral but are socio-technical and thus embedded with cultural, religious, and historical meanings that shape the communities they represent. Religious and cultural heritages have long histories of sharing and generating knowledge through folk tales, rituals, and other communal practices \cite{Zort2023}. For example, much of these traditions often depend not only on written texts but also on oral storytelling and embodied experiences to pass knowledge from one generation to the next. Lived experience thus plays a crucial role in shaping how stories and meanings are created, adapted, and maintained, as telling the story itself becomes a way of constructing meaning. As proposed by scholars of religion, artificial intelligence and machine learning, AI can serve as a tool to advance the study of religion \cite{yao2024role}, while also enabling religious frameworks to inform our understanding of AI itself \cite{reed2021ai}. 

\subsection{Case Study 4 : AI for Task Instruction}

Multi-modal systems integrate various input and output modalities (such as speech, touch, gesture, gaze, and haptics) to enable richer and more natural interaction between humans and machines. While traditionally measured through technical performance and usability metrics, the real-world impact of these systems often hinges on the lived experiences of users. The recent advent of LLMs has also combined vision and language tasks through the use of Vision-Language Models (VLMs)~\cite{nguyen2025install}. From setting the temperature of your fridge to building one yourself, the internet is filled with instructional videos. However, these videos are hard to follow because the system shown uses a different version of the model than the one currently in use. In such cases, it is generally easier to ask the AI system specific questions related to your model of the system or device. Using applications with multiple modalities, it will be helpful for the end user to have an image input augmented with the natural language input when asking questions specific to a task. 

\subsubsection{The Problem} 
Mohan's~\cite{mohan2019building} research is in the direction of apprenticeship learning, where the user can be taught to perform seemingly complicated tasks like changing the cartridge on a printer with near real-time instructions using computer vision-guided instructions. Wang et al.~\cite{wang2019towards} demonstrate similar solutions from a robotics perspective with a UR5 arm mounted on a scooter that reaches out for objects and places them in a bag when pointed to by the user, easily extendable to haptic feedback and speech instructions. Nguyen et al.~\cite{nguyen2025install} present a system for video question-answering that lets users search for specific questions ranging from technical to everyday instructional tasks, and responds with instructional video snippets for each step needed to perform a task. 

\subsubsection{What did the Study find} 
While these studies provide us with solutions to good instruction generation using VLMs, the proposed solutions faces the challenge of noise in language, speech and vision, in case the input is distorted with noise~\cite{malik2023extreme1, malik2023extreme2}. The main point of failure is lack of accounting for the mental models of task performance may differ user to user \cite{allen1997mental}. A visual overload of information and excessive decision-making demands can hinder overall understanding of the task \cite{sephton2013decision}. The solutions also pose privacy and security concerns in case of processing private and sensitive data in an open setting, say reading out banking information in a public space.  

\subsubsection{Role of Lived Experiences}  
As mentioned in the framework, understanding user mental models for learning tasks, the system could perform a context-aware modality switching to pre-process the input before sending it to the model, helping cut down the noise in the system. Contextual awareness also enables these systems to consider their operational environment and location, helping to ensure compliance with relevant privacy standards \cite{schaub2015context}. This dual benefit of cognitive alignment and contextual sensitivity enhances both system usability and trustworthiness. 

\section{Discussion}

Understanding and integrating human lived experiences in AI models is crucial for ensuring AI systems align with human values, needs, and well-being. Lived experiences shape how individuals interact with LLMs, influencing AI safety, design, and knowledge representation. Existing research highlights risks such as over-reliance \cite{passi2022overreliance}, compromised trust \cite{buccinca2021trust}, and psychological harm \cite{palka2023ai, chandra2024lived}, which are mediated by individual contexts \cite{kuper2025psychological}. These can be mitigated in part by inclusion of lived experiences. For instance, we find that personal experiences (and reflections) can play a pivotal role in defining the consumption of technology. For humans, cognition involves a complex interplay between external perceptions and internal explorations \cite{barsalou2014cognitive, mead1934mind, antony2001consciousness}. This work contributes to advancing that understanding by offering a clearer, more systematic approach—through a practical framework—for integrating lived experiences into AI development to foster better inclusivity. In this section, we discuss the broader implication of our work regarding importance of inclusion of pluralistic experiences, incorporating lived experiences with changing society and highlight the dynamic nature of the LEAF. Finally, we discuss the limitations of our work.

\subsection{Importance of inclusion of pluralistic experiences}

In recent years, interactions with technology have increasingly shaped how individuals perceive their sense of self and community. For instance, echo chambers formed by personalized algorithms as well as engagements with AI conversational agents can create the illusion of singular, enclosed realities \cite{belk2013extended, feige2019compensating}. The concept of lived experience, by contrast, is rooted in interpretiveness \cite{ellis1992investigating}. This interpretive nature enables multiple, coexisting perceptions of the same event or interaction. Within this plurality lies the unique richness of human perspective and creativity. By embedding such diversity into the design of AI tools, we not only center human involvement in technological systems but also enhance their effectiveness.

\subsection{Incorporating lived experiences in an ever-evolving world} Human experiences and expectations are not static, they are constantly changing in response to changes in society around us, and technological advancements~\cite{russon2003human, socialchangetheo, hbrTechnologyHuman}. At the same time, with the rapid development of AI and its increasing integration into human ecosystems, the way individuals interact with technology and perceive it is undergoing transformation~\cite{raees2024explainable, afroogh2024trust}. As a result, the notion of human lived experience in the age of AI is itself evolving. While our work presents dimensions of lived experience based on past research across various disciplines, it is likely that new AI-specific dimensions of human lived experiences may emerge in the future, something future work could explore. Moreover, as human-AI interactions become more frequent and complex with time, additional stages may be required and introduced in AI design and development pipeline to accommodate these new experiential factors.

\subsection{Dynamism within the LEAF} Beyond establishing the relevance and dimensions of lived experience, it is equally important to articulate a path forward. While comprehensive, our framework remains flexible, allowing new stages to be added or existing ones to be modified based on specific use cases or technological advancements. Similarly, other studies have emphasized that the integration of researchers’ evolving positionalities and lived experiences can critically influence not only the framing of research questions but also the interpretive lenses applied during system design and evaluation \cite{dembele2024researching, 10.1145/3715275.3732055}. This means inherent systems often suffer with lack of dynamism when those provisions are not provided. 



\subsection{Limitations and Future Work}

Finally, while this framework has been contextualized within domains such as healthcare, religion, education, and everyday life, its broader applicability remains to be discussed and validated. Future work should explore its relevance in additional sectors, such as finance, entertainment, legal systems, and more where lived experience may manifest differently. Advancing this work will also require the development of new tools and metrics capable of capturing the nuanced, multifaceted nature of lived experience. We envision future research extending this framework through domain-specific adaptations, innovative methodological approaches, and large-scale empirical validation.

\section{Conclusion}

This paper has underscored the critical role of lived experiences in shaping human-centered, contextually grounded, and ethically aligned AI systems. Through theoretical grounding and illustrative scenarios, we have shown that the exclusion of lived, subjective experiences from the AI design pipeline can lead to systems that misalign with users’ intentions, contexts, and values—resulting in miscommunication, harm, and systemic bias. Our discussion bridges gaps between technical design, social experience, and cultural specificity, advocating for methodological pluralism that includes first-person perspectives, embodied interactions, and contextual knowledge. We argue that integrating lived experiences is not merely an ethical imperative but also a design necessity—essential for closing the gulf between user intentions and system behavior, and for fostering trust, usability, and inclusivity. As AI systems increasingly mediate everyday decision-making and identity formation, this work calls for a sustained commitment to embedding the pluralities of human life into the design and development of intelligent technologies.

\bibliography{aaai25}

\end{document}